\documentclass[prb,preprint]{revtex4}%
\usepackage{amsmath}
\usepackage[dvips]{graphicx}%
\usepackage{amsfonts}%
\usepackage{amssymb}

\begin{document}

\title{An investigation into a half page from Newton's Principia in the wake of Chandra}
\author{Robert Golub}
\affiliation{Hahn Meitner Institut\\Glienckerstr. 100\\ 14109 Berlin, Germany}
\email{golub@hmi.de}
\author{W. M. Snow}
\affiliation{University Cyclotron Facility\\ 22401 Milo B
Sampson Lane\\ Bloomington, Indiana 47408\\ USA}
Subj-class: History of Physics

\begin{abstract}
There is a section in Chandrashekar's ''Newton's Principia for the Common
Reader '', (Clarendon Press, Oxford, 1995) in which he claims to find a small
error in the Principia. . However we believe that there is a mistake of
interpretation underlying Chandra's claim and that the Principia is correct as
it stands. This short paper describes  Chandra's misinterpretation of a
geometric construction of Newton and gives an outline of Newton's
demonstration by following the standard English version of the
\textit{Principia} line by line and converting it into modern mathematical
notation in the spirit of Chandra's book.

\end{abstract}

\maketitle
\volumeyear{year}
\volumenumber{number}
\issuenumber{number}
\eid{identifier}
\date{[Date text]date}
\received{date}
\revised{date}
\accepted{date}
\published{date}
\startpage{101}
\endpage{102}
\tableofcontents

\section*{Introduction}

Towards the end of his life the great astrophysicist Subrahmanyan
Chandrashekar wrote a very interesting, educational, and entertaining book
which was a reader's guide to Newton's Principia (''Newton's Principia for the
Common Reader '', Clarendon Press, Oxford, 1995). Chandra characterized the
nature of his project in the prologue as '' an undertaking by a practising
scientist to read and comprehend the intellectual achievement that the
\textit{Principia} is ''. The resulting book is a wonderful translation of
Newton's arguments into modern language and mathematical notation accompanied
by historical and physical commentary.

There is a section of Chandra's book in which he claims to find a small error
in the Principia. This is the sort of claim that naturally draws one's
attention. However we believe that there is a mistake of interpretation
underlying Chandra's claim and that the Principia is correct as it stands.
This short paper describes  Chandra's misinterpretation of a geometric
construction of Newton and gives an outline of Newton's demonstration by
following the standard English version of the \textit{Principia} line by line
and converting it into modern mathematical notation in the spirit of Chandra's book.

First a brief description of the issue, which concerns Newton's prescription
for determining the orbits under an inverse cube force (Proposition XLI,
corollary, III, page 132, section 50 in Chandrashekar) which appears about a
quarter of the way through the \textit{Principia}. After Newton has introduced
his laws of motion and derived Kepler's laws, he is in the midst of deriving
properties of orbits for various types of centripetal forces, employing such
ideas as energy conservation and the clear formulation of initial value
problems. In Corollary III to Proposition XLI concerning the orbits under a
centripetal force, Newton outlines a geometric construction for determining
the orbits under an inverse cube force. This construction relies on the use of
an auxiliary conic section, the curves VRS in the figure shown below.
According to Newton's construction when the auxiliary curve is a hyperbola,
the constructed orbit (the curves VPQ) spirals in towards the center and when
the auxiliary curve is an ellipse the constructed orbit is a hyperbola. This
introduction of an auxiliary curve has led to some confusion and led
Chandrashekar to assert, in section 50 of his book, that the correct result is
the other way around and that the statement in the \textit{Principia} must be
a misprint. It seems that the introduction of an auxiliary curve led
Chandrashekar to a slight misinterpretation of the argument which it is the
purpose of this note to clarify.

In this paper all ''quotes'' from the \textit{Principia}, which was of course
originally written in Latin, are from the English translation ''Sir Isaac
Newton's Mathematical Principles of Natural Philosophy and His System of the
World'', translated by A. Motte, revised by F Cajori, University of California
Press, Berkeley, 1934). \cite{newprinc}%

\begin{center}
\includegraphics[bb=100 300 470 560,
height=4.0511in,
width=4.2454in
]%
{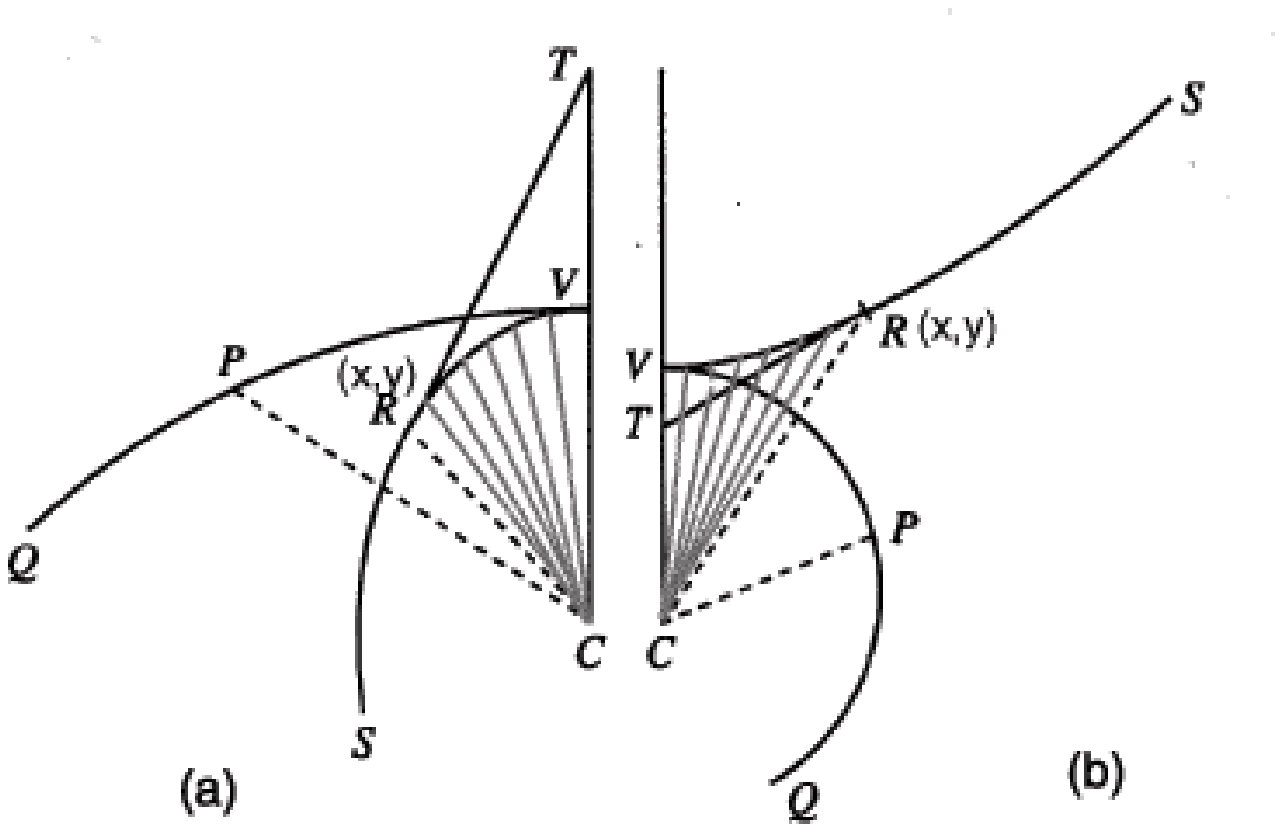}
\\
Fig. 1) The curve VRS is tha auxiliary conic section used by Newton to
construct the orbit VPQ. In fig. (a) the auxiliary curve is an ellipse leading
to an orbit, whose radius is given by CT (RT is the tangent to VRS) which
grows continuously as the auxiliary point R(x,y) moves down the ellipse to S.
In fig. (b) the auxiliary curve VRS is a hyperbola leading to an orbit VPQ
whose radius (CT) decreases as the auxiliary point R(x,y) moves towards S.
\ \ The area VRC proportional to the angle along the orbit, VCP, is
indicated.\ \ See text below. (Adapted from Chandrashekar)
\end{center}

In his discussion \ Chandra quotes the following passage from the Principia:

\begin{quote}
''..therefore if the conic section VRS be a \underline{hyperbola}, the body
will descend to the centre (\emph{along VPQ}); but if it be an \underline
{ellipse}, it will ascend continually and go farther and farther \emph{in
infinitum}. And on the contrary. (Parentheses added)
\end{quote}

Chandra then comments (page 180)

\begin{quote}
''On the face of it, one would conclude that the words 'hyperbola' and
'ellipse' (underlined) have been interchanged by a simple oversight.
Certainly, an orbit which is a hyperbola in the ($r,t$) plane ascends to
infinity while $\phi$ tends to a finite limit, while an orbit which is an
ellipse on the ($r,t$) plane descends to the centre in a spiral orbit...''
\end{quote}

In this statement and in the discussion on p. 176 it is clear that Chandra is
interpreting the diagrams given by Newton as being orbits in the $\left(
r,t\right)  $ plane but Newton never mentions time in his discussion and we
will see that if one interprets the curves VRS as auxiliary curves and VPQ as
the orbits the statement in the Prinicpia is correct as it stands.

At the beginning of section 50. Chandra states the word ''body'' is not used
with its standard meaning. Newton states

\begin{quote}
''..from the place V there sets out a body with a just velocity...that body
will proceed in a curve VPQ ...'', (page 132)
\end{quote}

(see complete quote below). However if we accept VPQ as the trajectory, then
there does not appear to be anything wrong with this use of the word ''body''.

In the next section we repeat Chandra's calculation of the orbit and in
section 3 we show that following Newton's prescription leads to the same orbit.

\section{Orbits for an inverse cube force}

Following Chandra (equation 1, p. 174) we write the conservation of energy
\begin{align}
\dot{r}^{2}  &  =\frac{1}{r^{2}}-\frac{h^{2}}{r^{2}}+C=\frac{(1-h^{2})}{r^{2}%
}+C\label{1}\\
&  =(1-h^{2})\left(  \frac{1}{r^{2}}-\frac{1}{\beta^{2}}\right) \nonumber
\end{align}
where the first term is the potential for the $1/r^{3}$ force and the second
term is the centrifugal potential. (We took $\dot{r}=0$ at $r=\beta)$.
Conservation of angular momentum yields%
\begin{equation}
\frac{d\phi}{dt}=\frac{h}{r^{2}} \label{2}%
\end{equation}

Considering first $h^{2}>1$ (\ref{1}) and the initial condition yields%
\begin{align}
\frac{dt}{dr}  &  =\frac{\beta r}{\sqrt{h^{2}-1}\sqrt{r^{2}-\beta^{2}}%
}\label{3b}\\
\frac{d\phi}{dr}  &  =\frac{d\phi}{dt}\frac{dt}{dr}=\frac{h\beta}{r\sqrt
{h^{2}-1}\sqrt{r^{2}-\beta^{2}}} \label{3a}%
\end{align}

\bigskip Evaluating the integral (\ref{3a}) we find (taking $\phi=0$ at
$r=\beta$)%
\begin{align}
\phi &  =\frac{h}{\sqrt{h^{2}-1}}\cos^{-1}\frac{\beta}{r}\nonumber\\
\frac{\beta}{r}  &  =\cos(\frac{\sqrt{h^{2}-1}}{h}\phi) \label{4}%
\end{align}
which is the equation of a hyperbola (Chandrashekar equ. 25, p 178). Using
conservation of angular momentum (\ref{2}) or evaluating the integral
(\ref{3b}) we find that the orbit in the $\left(  r,t\right)  $ plane is
indeed a hyperbola%
\begin{align}
\frac{r^{2}}{\beta^{2}}-\frac{t^{2}}{\alpha^{2}}  &  =1\label{3aa}\\
\alpha^{2}  &  =\frac{\beta^{2}}{h^{2}-1}\nonumber
\end{align}
as given by Chandra.

\bigskip

In the case of $h^{2}<1$ it is easy to show that
\begin{equation}
\frac{\beta}{r}=\cosh(\frac{\sqrt{1-h^{2}}}{h}\phi) \label{4a}%
\end{equation}
and
\begin{equation}
\frac{r^{2}}{\beta^{2}}+\frac{t^{2}}{\alpha^{2}}=1 \label{4aa}%
\end{equation}

Of course Chandra is correct when he says the orbit of equations (\ref{3aa})
is a hyperbola in the ($r,t$) plane and that of equations (\ref{4aa}) is an
ellipse in that plane but that does not form any part of Newton's argument. In
the next section we look directly at Newton's suggested construction.

\section{ Corollary III; Newton's geometrical construction of the orbits for
an inverse cube force}

In the following we will calculate the orbits following Newton's prescription
step by step. The references are to the figure The quoted passages are from
p.132 Principia, 3rd ed, etc....

\begin{quotation}
(i) \emph{''If to the centre C and the principal vertex V, there be described
a conic section VRS;...''}
\end{quotation}

Designating the point $R$ by the coordinates $\left(  x,y\right)  $ with $y$
axis along CV and $x$ axis directed to the right from the origin C, we can
write the equations for the conic section taken as an ellipse (lhs fig)%
\begin{equation}
\frac{y^{2}}{\beta^{2}}+\frac{x^{2}}{\alpha^{2}}=1 \label{5}%
\end{equation}
or, alternately as a hyperbola (rhs fig)%
\begin{equation}
\frac{y^{2}}{\beta^{2}}-\frac{x^{2}}{\alpha^{2}}=1 \label{6}%
\end{equation}

\begin{quote}
(ii) \emph{''...and from any point thereof, as R, there be drawn the tangent
RT meeting the axis CV indefinitely produced in the point T;...''}
\end{quote}

From (\ref{5}) and (\ref{6}) we calculate%
\[
\frac{dy}{dx}=\mp\frac{x}{y}\frac{\beta^{2}}{\alpha^{2}}
\]
with the top (bottom) sign applying to the ellipse (\ref{5}) (hyperbola
\ref{6}). Then the distance CT is given by%
\begin{align*}
CT  &  =y-x\frac{dy}{dx}=y\pm\frac{\beta^{2}}{y}\frac{x^{2}}{\alpha^{2}}\\
&  =y\pm\frac{\beta^{2}}{y}\left(  \pm\right)  \left(  1-\frac{y^{2}}%
{\beta^{2}}\right) \\
&  =\frac{\beta^{2}}{y}%
\end{align*}

\begin{quote}
(iii) \emph{''...and then joining CR there be drawn the right }(meaning
straight not perpendicular)\emph{ line CP, equal to the abscissa CT,...''}

The radius relative to the origin C of the point P is:%
\begin{equation}
r_{P}=CT=\frac{\beta^{2}}{y} \label{7}%
\end{equation}
\ \ \ \ \ 

\ \ \ \ \ \ (iv)\ \emph{''..making an angle VCP proportional to the sector
(area) VCR;...''}
\end{quote}

Thus%
\[
\phi\sim A_{VCR}=\int ydx-\frac{xy}{2}
\]
\emph{\ }\ \ where $\frac{xy}{2}$ is the area of the triangle formed by CR,
the x axis, and a vertical line through R.%
\[
A_{VCR}=\frac{\beta}{\alpha}\int_{0}^{x}\sqrt{\alpha^{2}\mp x^{2}}dx-\frac
{x}{2}\frac{\beta}{\alpha}\sqrt{\alpha^{2}\mp x^{2}}
\]

\bigskip For the hyperbola (lower sign, \ref{6}) we have%
\begin{align*}
A_{VCR}  &  =\frac{\beta}{\alpha}\int_{0}^{x}\sqrt{\alpha^{2}+x^{2}}%
dx-\frac{x}{2}\frac{\beta}{\alpha}\sqrt{\alpha^{2}+x^{2}}\\
&  =\frac{\beta}{\alpha}\left(  \allowbreak\frac{1}{2}x\sqrt{\left(
\alpha^{2}+x^{2}\right)  }+\frac{1}{2}\alpha^{2}\left(  \sinh^{-1}\frac
{x}{\alpha}\right)  \right)  -\frac{x}{2}\frac{\beta}{\alpha}\sqrt{\alpha
^{2}+x^{2}}\\
&  =\frac{1}{2}\beta\alpha\left(  \sinh^{-1}\frac{x}{\alpha}\right)  =\frac
{1}{2}\beta\alpha\sinh^{-1}\sqrt{\frac{y^{2}}{\beta^{2}}-1}\\
&  =\frac{1}{2}\beta\alpha\cosh^{-1}\frac{y}{\beta}%
\end{align*}

\bigskip where we used (\ref{6}). Thus%
\[
\frac{y}{\beta}=\cosh K^{\prime}\phi
\]
and%
\[
r_{P}=\frac{\beta}{\cosh K^{\prime}\phi}
\]
which is equation (\ref{4a}). Thus we see that the curve generated by P
following Newton's prescription (quotes (iii) and (iv) above) for the conic
section VRS being an hyperbola (\ref{6}) does indeed spiral into the center.

\begin{quotation}
\emph{''..and if a centripetal force inversely proportional to the cubes of
the distances of the places from the centre, tends to the centre C: and from
the place V there sets out a body with a just velocity in the direction of a
line perpendicular to the right (straight) line CV; that body will proceed in
a curve VPQ , which the point P will always touch; \underline{\emph{and
therefore if the conic section VRS be an hyperbola}}, \underline{\emph{the
body will descend to the centre,}} \ }
\end{quotation}

as the point R moves up the curve VRS the point T moves toward the origin.

Further for VCR taken as an ellipse (upper sign, \ref{5}) we have%
\begin{align*}
A_{VCR}  &  =\frac{\beta}{\alpha}\int\sqrt{\alpha^{2}-x^{2}}dx-\frac{x}%
{2}\frac{\beta}{\alpha}\sqrt{\alpha^{2}-x^{2}}\\
&  =\frac{\beta}{\alpha}\left(  \frac{1}{2}x\sqrt{\left(  \alpha^{2}%
-x^{2}\right)  }+\frac{1}{2}\alpha^{2}\arcsin\frac{x}{\alpha}\right)
-\frac{x}{2}\frac{\beta}{\alpha}\sqrt{\alpha^{2}-x^{2}}\\
&  =\frac{1}{2}\beta\alpha\arcsin\frac{x}{\alpha}%
\end{align*}
Then
\begin{align*}
\phi &  =KA_{VCR}=K\frac{1}{2}\beta\alpha\arcsin\frac{x}{\alpha}=\frac{1}%
{2}\beta\alpha\arcsin\sqrt{1-\frac{y^{2}}{\beta^{2}}}\\
&  =\frac{K}{2}\beta\alpha\arccos\frac{y}{\beta}%
\end{align*}
or%
\[
\frac{y}{\beta}=\cos K^{\prime}\phi
\]
and (\ref{7})%
\[
r_{P}=\frac{\beta^{2}}{y}=\frac{\beta}{\cos K^{\prime}\phi}
\]
which is equation (\ref{4}). Thus we see that the curve generated by P
following Newton's prescription for the conic section VRS being an ellipse
(\ref{5}) \ is indeed a hyperbola;

\begin{quote}
\emph{''...but if it (VRS) be an ellipse, it (the body) will ascend
continually, and go further off }in infinitum.\emph{'' }
\end{quote}

\emph{\ }as the point R moves down the ellipse VRS and the tangent approaches
the vertical, the point T moves off to infinity.

\bigskip

\section{Conclusion}

We have shown that following Newton's geometric prescription one generates the
correct orbits for the inverse cube force; taking the auxiliary curve as a
hyperbola leading to orbits that spiral in towards the center while an ellipse
as auxiliary curve leads to hyperbolic orbits flying out to infinity so that
there is no confusion in Newton's presentation.

To return to the quote from Newton cited by Chandra (see introduction)

\begin{quote}
''..therefore if the conic section VRS be a \underline{hyperbola}, the body
will descend to the centre (\emph{along VPQ}); but if it be an \underline
{ellipse}, it will ascend continually and go farther and farther \emph{in
infinitum}. And on the contrary.'' (Parentheses added)
\end{quote}

to which Chandra had the following comment:

\begin{quote}
''On the face of it, one would conclude that the words 'hyperbola' and
'ellipse' (underlined) have been interchanged by a simple oversight.
Certainly, an orbit which is a hyperbola in the ($r,t$) plane ascends to
infinity while $\phi$ tends to a finite limit, while an orbit which is an
ellipse on the ($r,t$) plane descends to the centre in a spiral orbit...''
\end{quote}

We see that Newton's statement was indeed correct and can see, as well, the
origin of Chandra's statement.

Certainly none of this is meant as a criticism of Chandra, who was clearly a
very creative and insightful scientist. (The book in question was published in
the year of his death at age 85). We feel that Chandra would not have wanted
such an apparent error of interpretation in his wonderful work to stand.

For us the interesting thing was the fun of following Chandra's lead in
reading a small piece of one of Newton's original arguments in the Principia
and reconstructing the reasoning for ourselves. Although we verify that
Newton's construction is correct, the really interesting question is how he
came up with the idea for such a construction in the first place. It is also
interesting to note how succinctly Newton was able to present the argument,
requiring about half a page including the figure. Newton himself explains his reasoning:

\begin{quote}
''\emph{All these things follow from the foregoing Proposition }(XLI, page
139, which, as explained by Chandra, is an exposition of the general energy
integral method for solving the motion of a particle under an arbitrary
central force), \emph{by the quadrature of a certain curve, the invention of
which as being easy enough, for brevity's sake I omit.''}
\end{quote}

This passage was quoted by Chandra (page 174) as an introduction to his
solution of the orbits using the energy integral method in which he identifies
the curves VRS with the orbit in the $(r,t)$ plane, something, which seems not
to have been intended by Newton.

The \textit{Principia} has a reputation as one of the least frequently-read
''great books''. For the usual reasons of time pressure felt so strongly in
science education curricula, few instructors try to teach the subject of
mechanics by guiding students through Newton's derivations in the Principia.
However with the aid of Chandra's lovely book and similar books \cite{a},
\cite{b} it is now far more practical for an instructor to consider using a
few selected excerpts from the \textit{Principia} to give students a taste of
the real Newton. We feel that some students would be excited by the
opportunity to confront a piece of the \textit{Principia} with some guidance
and we hope that more instructors will consider this possibility.

\end{document}